\documentclass[aps,pra,twocolumn,amsmath,amssymb,floatfix]{revtex4}
\usepackage{graphicx}
\usepackage{float}
\usepackage[dvips]{color}

\def\bc{\begin{center}}
\def\ec{\end{center}}

\def\beq{\begin{equation}}
\def\eeq{\end{equation}}

\def\d{\downarrow}
\def\u{\uparrow}

\begin{document}
\title{Noise correlations of a strongly attractive spin-1/2 Fermi gas in an optical lattice}
\author{O. Fialko and K. Ziegler}
\affiliation{Universit\"at Augsburg, Germany}

\begin{abstract} 
We calculate density-density correlations of an expanding gas of strongly attractive ultra-cold spin-1/2 fermions
in an optical lattice. The phase diagram of the tightly bound fermion pairs exhibits a Bose-Einstein condensed state 
and a Mott insulating state with a single molecule per lattice site. We study the effects of quantum 
fluctuations on the correlations in both phases and show that they are especially important in the Bose-Einstein 
condensate state, leading to the appearance of singular peaks. In the Mott insulating state the correlations are 
characterized by sharp dips. This can be utilized in experiments to distinguish between these two phases.  
\end{abstract}

\maketitle

\section{Introduction}
Experiments with ultra-cold atoms have opened many directions to study ground state properties of complex
many-body systems, such as Bose-Einstein condensation and fermion pairing. 
The introduction of optical lattices has added the possibility of creating new ground states \cite{bloch08}. 
A prominent example is the formation of a Mott insulating (MI) phase for strongly interacting bosons \cite{greiner02} 
or fermions \cite{esslinger08}. 

Another interesting direction was brought to the field by the unprecedented control over the interaction between atoms 
via a magnetic Feshbach resonance \cite{stoof04}. By means of the latter it became possible to turn fermions
with two hyperfine internal states into bosons \cite{melo08} by pairing. For weak attraction the 
Bardeen-Cooper-Schrieffer (BCS) theory predicts
an instability towards formation of large Cooper pairs, whose size can be much larger than the
separation between atoms in the Fermi gas. If the attraction is turned on stronger, then these Cooper 
pairs become smaller 
and eventually, for very strong interaction, they behave as point-like (hard-core) bosonic molecules.
Although there is no sharp distinction between the BCS state and a condensate of local pairs of fermions,
the cross-over from the weakly interacting to the strongly interacting Fermi gas is of substantial interest 
\cite{rink84}. The experimental creation of bosonic molecules
in a Fermi gas atoms as well as their Bose-Einstein condensation has been recently reported \cite{jin03}.
This BCS-BEC crossover has also been realized in an optical lattice \cite{esslinger05}. In the presence of an
optical lattice the tightly bound molecules in the BEC regime may undergo a 
quantum phase transition from a BEC state to a MI state with one particle per 
lattice site. This leaves us with question how to identify the two phases by simple measurements.
 
A clear signature of superfluidity in a system of bosons is the presence of sharp density peaks due to phase
coherence in time-of-flight 
experiments \cite{greiner02}. In the MI phase, on the other hand, the density peaks are washed out
due to the absence of phase coherence (see \cite{kato08}). To characterize this phase it is useful to study
the noise correlations in the Fermi gas \cite{theory}, where a characteristic feature is a 
dip that has been seen experimentally \cite{bloch06}.  

In this paper, we study density-density correlations of an expanding gas of ultra-cold spin-1/2 fermions
realized in an optical lattice.  
We discuss that the density-density correlations reveal the fermionic nature of the gas in the MI state 
and its bosonic nature in the Bose-Einstein condensate (BEC). We show that the effects of quantum fluctuations on the 
correlations are important. 
 

\section{Model}

An effective Hamiltonian for fermions in an optical lattice 
near a Feshbach resonance was derived in Ref. 
\cite{duan05}. Under typical experimental conditions, the resulting Hamiltonian can be mapped to an effective 
single-band model, which comprises the local dressed bosonic molecules and individual fermionic atoms separately. 
In this paper we consider a purely fermion model in an optical lattice
with attractive interaction and with two internal states, denoted by the pseudospin projections 
$\uparrow$ and $\downarrow$, respectively. 
The fermionic field operator $\hat{\psi}_{\sigma}({\bf r})$ is expanded in a basis of Wannier 
functions \cite{kohn59} as
\beq
\hat{\psi}_{\sigma}({\bf r})=\sum_{i,n}w_n ({\bf r}-{\bf R}_i)\hat{c}_{i,n,\sigma}.
\label{wannier}
\eeq 
which allows us to use the tight-binding representation.
Here, $\hat{c}_{i,n,\sigma}$ is the annihilation operator for particles in the Wannier state $n$ at lattice 
site ${\bf R}_i$. For a deep optical lattice only the lowest Wannier state is taken into account with 
$\hat{c}_{i,\sigma}\equiv\hat{c}_{i,0,\sigma}$. In this case the lattice Hamiltonian reads
\[
\hat{H}=-\frac{\bar{t}}{2d}\sum_{\sigma=\uparrow,\downarrow}\sum_{\langle
i,j\rangle}\hat{c}_{i\sigma}^{\dagger}\hat{c}_{j\sigma}
-\frac{J}{2d}\sum_{\langle
i,j\rangle}\hat{c}_{i\uparrow}^{\dagger}\hat{c}_{j\uparrow}\hat{c}_{i
\downarrow}
^{\dagger}\hat{c}_{j\downarrow}
\]
\beq
- \sum_{\sigma=\uparrow,\downarrow}\sum_{i}\mu_\sigma 
\hat{c}_{i\sigma}^{\dagger}\hat{c}_{i\sigma}.
\label{model}
\end{equation}
Nearest-neighbor tunneling of the individual fermions is described by the parameter $\bar{t}$.
Moreover, there is also a nearest-neighbor 
attraction with parameter $J$ which can be understood as a tunneling term of fermionic pairs (molecular tunneling).
A similar model of a 
mixture of hard-core bosons and fermions in a lattice was discussed in Ref. \cite{micnas07}. However, in our
model the hard-core bosons $\hat{b}_{i}$ are given by pairs of fermion operators 
$\hat{c}_{i\uparrow}\hat{c}_{i\downarrow}$. This means that the fermions can transmute 
dynamically into hard-core bosons by local pairing.

\section{Phase diagram}

We apply the functional integral formalism developed for the Hamiltonian in Eq. (\ref{model})
in \cite{ziegler,we}. 
The partition function of a grand-canonical ensemble of fermions with chemical potential
$\mu$ at inverse temperature $\beta$
can be expressed in terms of an integral over a fermionic (Grassmann) field $\psi$ and complex 
molecular fields $\phi,\chi$ as
\begin{equation}
Z=\int e^{-\int_0^{\beta} d\tau {\cal L}}{\cal{D}}[\psi,\phi,\chi]
\ 
\end{equation}
with the Lagrangian 
\beq
{\cal L}=\sum_{i,j}
\bar{\phi}_{i}\hat{v}^{-1}_{i,j}\phi_{j}
+\frac{1}{2J}\sum_{i}\bar{\chi}_{i}\chi_{i}-\sum_{i,j} 
\hat{{\cal{C}}}^{\dagger}_{i} \hat{\bf{G}}^{-1}_{ij} 
\hat{{\cal{C}}}_{j}
\ ,
\label{action}
\eeq
where we have used the notation 
$\hat{{\cal{C}}}^{\dagger}=(\psi_{\uparrow},\bar{\psi}_{\downarrow})$ 
and $\hat{{\cal{C}}}=(\psi_{\downarrow},\bar{\psi}_{\uparrow})^T$, and
with the inverse fermionic Green's matrix 
\beq
\hat{\bf{G}}^{-1}=\left(\begin{array}{cc} -i\phi-\chi &\partial_{\tau}+\mu+\bar{t}  \\
\partial_{\tau}-\mu-\bar{t} & i\bar{\phi}+\bar{\chi}  \end{array} \right).
\label{greensf}
\eeq
Here $\hat{v}_{i,j}=\frac{J}{2d}\delta_{|i-j|,1}$.
$\hat{t}$ is the nearest-neighbor tunneling matrix with elements $\bar{t}/2d$.

\begin{figure}[t]
\begin{center}
\includegraphics[width=8cm]{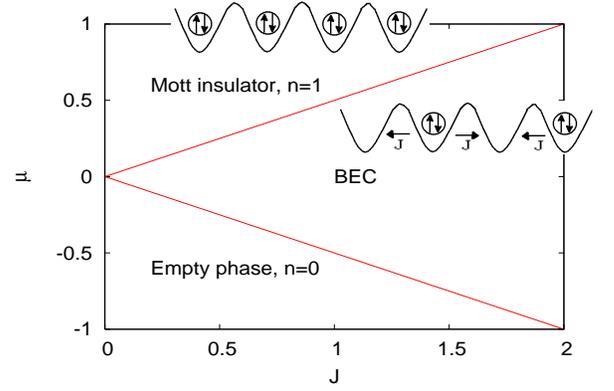}
\end{center}
\caption{Phase diagram for $\bar{t}=0$ and $k_B T=0$. 
The three phases are the BEC with a non-vanishing condensed density, the MI 
states with vanishing condensed density and with one molecules per lattice site and the empty phase.
$\mu$ and $J$ are in arbitrary energy units.}
\label{diagram}
\end{figure}

The integration over the complex molecular fields $\phi,\chi$ can be performed in saddle-point 
approximation \cite{ziegler,we}. For a vanishing fermionic tunneling, i.e. by assuming a situation where
all fermions are paired up, 
the phase diagram is depicted in Fig. \ref{diagram}. There are three phases: the BEC
of molecules with the condensed fraction $n_0=(J^2-4\mu^2)/(4J^2)$, the MI state with one particle per 
site and $n_0=0$ and the empty phase. 
Fluctuations around the saddle point provide the low-energy excitations of the bosonic molecules. They are gapless 
in the BEC phase:
\beq
\epsilon_{{\bf k}}=\sqrt{4J^2g_{{\bf k}}n_0 +4g_{{\bf k}}^2\mu^2},
\label{spectrum0}
\eeq
where $g_{{\bf k}}=1-1/d \sum_{i=1}^d \cos k_i$ is the dispersion of the free Bose gas. 
On the other hand, the excitations of the MI state have a gap $\Delta=2\mu-J>0$ :
\beq
\epsilon_{{\bf k}}=\Delta+Jg_{{\bf k}}
\ .
\eeq

\section{Density-density correlations}

In a typical experiment with ultra-cold atoms it is not easy to identify the spin projection.
Therefore, the detected fermion density should be described by a superposition of both spin projections.
For this purpose we introduce the spin-independent fermionic density
$\hat{n}=\hat{n}_{\uparrow}+\hat{n}_{\downarrow}$.
Then the truncated density-density correlation function in a time-of-flight experiment is given under the
assumption that the atoms expand freely after they have been released from the optical
lattice as \cite{zwerger08}
\[
C_{{\bf r},{\bf r}^{\prime}}\equiv\langle \hat{n}({\bf r})\hat{n}({\bf r}^{\prime})\rangle 
-\langle \hat{n}({\bf r})\rangle \langle\hat{n}({\bf r}^{\prime})\rangle
=
\]
\beq
\left(\frac{M}{\hbar 
t}\right)^{2d}\langle \hat{n}({\bf k})\hat{n}({\bf k}^{\prime})\rangle - \langle \hat{n}({\bf 
r})\rangle \langle\hat{n}({\bf 
r}^{\prime})\rangle. 
\label{coor}
\eeq
Here ${\bf k}$ is related to ${\bf r}$ by ${\bf k}=M{\bf r}/\hbar t$ with the effective mass $M/\hbar\propto 1/J$.
Free expansion assumes that atoms evolve independently. This can be achieved by switching off an optical 
lattice as well as by switching the magnetic field to values far from the strongly attractive regime suddenly 
before the expansion \cite{theory, bucker}. 
The above formula reveals then that the density distribution, observed after a fixed time-of-flight, is 
related to the momentum distribution of the particles trapped in the lattice before the expansion.
It is widely used to analyze ground state properties of ultra-cold atoms trapped by an optical lattice 
\cite{imambekov}.


Now we use the expansion of Eq. (\ref{wannier}) to express the density-density correlation of Eq. (\ref{coor}) 
in terms of the coordinates of the underlying optical lattice as
\[
C_{{\bf r},{\bf r}^{\prime}}=\left(\frac{M}{\hbar t}\right)^{2d}
|\tilde{w}({\bf k})|^2|\tilde{w}({\bf k}^{\prime})|^2\times
\]
\[
\sum_{i,i',j,j'}e^{i{\bf k}({\bf R}_i-{\bf R}_{i'})+i
{\bf k}^{\prime}({\bf R}_j-{\bf R}_{j'})}\langle
\hat{c}_{i\alpha}^{\dagger} \hat{c}_{j\beta}^{\dagger}\hat{c}_{j'\beta} \hat{c}_{i'\alpha}
\rangle,
\]
\beq
+\langle \hat{n}({\bf r})\rangle \delta({\bf r}-{\bf r}')-\langle \hat{n}({\bf
r})\rangle \langle\hat{n}({\bf r}^{\prime})\rangle
\label{noise}
\eeq
where $\tilde{w}$ is a Fourier transform of the Wannier function.
Using the saddle-point approximation of the complex fields ($\phi,\chi$) we can evaluate
$\langle\hat{c}_{i\alpha}^{\dagger} \hat{c}_{j\beta}^{\dagger}\hat{c}_{j'\beta} \hat{c}_{i'\alpha}
\rangle$. Since the fermion field 
appears only in a quadratic form in Eq. (\ref{action}), the integration are
given by a Wick contraction of pairs of fermions:
\[
\langle \hat{c}_{i\alpha}^{\dagger} \hat{c}_{j\beta}^{\dagger}\hat{c}_{j'\beta} \hat{c}_{i'\alpha}
\rangle=
\langle \hat{c}_{i\alpha}^{\dagger}\hat{c}_{i'\alpha}\rangle
\langle \hat{c}_{j\beta}^{\dagger}\hat{c}_{j'\beta}\rangle
-\langle \hat{c}_{i\alpha}^{\dagger} \hat{c}_{j'\beta}
\rangle \langle \hat{c}_{j\beta}^{\dagger} \hat{c}_{i'\alpha}
\rangle+
\]
\beq
\langle \hat{c}_{i\alpha}^{\dagger} \hat{c}_{j\beta}^{\dagger}\rangle\langle\hat{c}_{j'\beta} \hat{c}_{i'\alpha}
\rangle
\label{contractions}
\eeq  
The fermionic expectation values on the right-hand side
are matrix elements of the fermionic Green's function $\hat{\bf{G}}$ of Eq. (\ref{greensf}):
\beq
\langle \hat{c}_{i\uparrow}^{\dagger} \hat{c}_{j\uparrow}\rangle = \langle \hat{c}_{i\downarrow}^{\dagger} 
\hat{c}_{j\downarrow}\rangle =\frac{1}{\beta}\sum_n {\bf
G}_{ij}^{12}(\omega_n)
\label{nf}
\eeq
and
\beq
\langle \hat{c}_{i\downarrow}^{\dagger}
\hat{c}_{j\uparrow}^{\dagger}\rangle=\frac{1}{\beta}\sum_n
{\bf G}_{{ij}}^{11}(\omega_n), \ \
\langle \hat{c}_{i\uparrow}\hat{c}_{j\downarrow}\rangle=\frac{1}{\beta}\sum_n {\bf
G}_{ij}^{22}(\omega_n),
\label{n0}
\eeq
where we have summed over Matsubara frequencies $\omega_n$. 
These expressions include also quantum fluctuations of the bosonic molecules
when we take into account Gaussian fluctuations around the saddle-point solutions. 
For instance, we can integrate over these fluctuations to evaluate the correlation function
\beq
\langle
\hat{c}_{i\alpha}^{\dagger} \hat{c}_{j\beta}^{\dagger}\hat{c}_{j'\beta} \hat{c}_{i'\alpha}
\rangle - \langle
\hat{c}_{i\alpha}^{\dagger} \hat{c}_{j\beta}^{\dagger}\rangle \langle\hat{c}_{j'\beta} \hat{c}_{i'\alpha}
\rangle.
\eeq

In general, the Green's function $\hat{\bf{G}}_{ij}$ is not diagonal in space due to the presence of the 
fermionic tunneling $\bar{t}$ but the decay is exponential with distance $|{\bf R}_i-{\bf R}_j|$. 
The exponential behavior can be approximated by taking only the diagonal part into account. 
In other words, in a strongly interacting Fermi gas the formation of bosonic molecules and their dynamics is the
dominant feature, and the individual tunneling of fermions is negligible.
Moreover, the Green's function depends on the fields $\phi$ and $\chi$.
In that case the density-density correlations reduces to
\[
C_{{\bf r},0} = \langle \hat{n}({\bf r})\rangle \delta({\bf r})+ 2\left(\frac{M}{\hbar t}\right)^{2d}
|\tilde{w}({\bf k})|^2|\tilde{w}(0)|^2\times
\]
\beq
\left[-\left|\sum_{i} 
e^{i{\bf k}{\bf R}_i}\langle \hat{c}^{\dagger}_{i\uparrow} \hat{c}_{i\uparrow}\rangle\right|^2 
+\left|\sum_{i} 
e^{i{\bf k} {\bf R}_i}\langle \hat{c}_{i\uparrow} \hat{c}_{i\downarrow}\rangle\right|^2
+{\cal{S}}({\bf k})\right],
\label{noise2}
\eeq
The first two terms constitute the density-density correlation function for fermions \cite{zwerger08},
whereas the third and the fourth terms account for the presence of the condensed molecules. In particular,
the fourth term describes the effect of quantum fluctuations and can be expressed as a
momentum distribution of the molecules \cite{we}
\beq
{\cal S}({\bf k})=\sum_{i,j}e^{{\bf k}({\bf R}_i-{\bf R}_j)}(\langle \hat{b}^{\dagger}_i \hat{b}_j\rangle
-\langle \hat{b}^{\dagger}_i\rangle \langle\hat{b}_j\rangle)
\eeq
with $\hat{b}^{\dagger}_i=\hat{c}_{i\d}^{\dagger} \hat{c}_{i\u}^{\dagger}$.
Phase fluctuations can destroy these terms, for instance, in the case of a MI state of the 
molecules. It is important to notice
that fermionic terms contribute with a negative sign in Eq. (\ref{noise2}), 
in contrast to the phase-coherent molecules, which contribute with
a positive sign. This indicates a competition of the fermionic and the molecular contribution to the 
density-density correlation function. 
This provides a concept for measuring the properties of a strongly interacting Fermi gas.

\section{Results}

The expressions $\langle \hat{c}^{\dagger}_{i\uparrow} \hat{c}_{i\uparrow}\rangle$ 
and 
$\langle \hat{c}_{i\uparrow} \hat{c}_{i\downarrow}\rangle$ are constant due to translational invariance:
\beq
C_{{\bf r},0}\propto N^2\delta_{{\bf k},{\bf G}}\left(-1+\frac{N_0\cal N}{N^2}\right)+{\cal{S}}({\bf k}),
\label{homog2}
\eeq
where we have denoted $2N=2\sum_i \langle \hat{c}^{\dagger}_{i\uparrow} \hat{c}_{i\uparrow}\rangle $ as the total 
number of fermions and $N_0{\cal N}=|\sum_{i}\langle \hat{c}_{i\uparrow} \hat{c}_{i\downarrow} \rangle|^2\propto 
{\cal{N}}^2|i\phi+\chi|^2$. ${\cal{N}}$ is the number of lattice sites. The ratio $n_0=N_0/N$ is the condensate 
fraction, i.e. the relative contribution of condensed molecules.

\begin{figure}[t]
\begin{center}
\includegraphics[width=6cm]{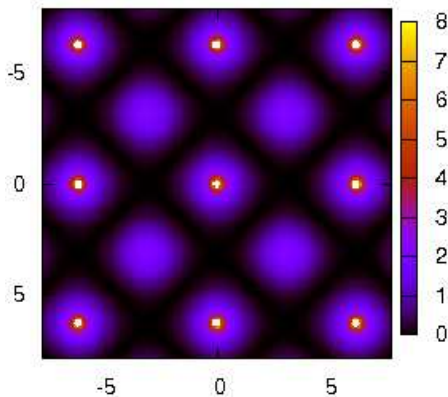}
\end{center}
\caption{Density-density correlation function of a strongly attractive spin-1/2 Fermi gas
in an optical lattice with $J=1$, $\mu=0.4$, $k_B T=0$, where the gas forms a BEC.
The singular behavior at the positions corresponding to the reciprocal lattice vectors is due to long range phase
coherence of the condensed molecules. There is also small modulation due to the tunneling of molecules. 
The axes are given in units of $\hbar t/Ma$, where $a$ is a lattice spacing.}
\label{BEC}
\end{figure}

{\bf BEC.}-- For momenta close to the reciprocal lattice vectors ${\bf G}$ 
the main contribution in the BEC comes from the term ${\cal{S}}({\bf k})$, 
since it is singular for $\epsilon_{\bf k}\sim0$: 
\beq
C_{{\bf r},0}\propto {\cal{S}}({\bf k})\approx\frac{4Jn_0+Jg_{\bf k}}{\epsilon_{\bf k}}(1-2 g_{\bf k})
\ .
\label{eq1}
\eeq
This results connects the density-density correlation function with the spectrum of the
molecular condensate given in Eq. (\ref{spectrum0}).
The infrared divergence of the momentum distribution is a general property of a BEC at 
$T=0$ \cite{pitaevskii}. The related 2D plot of the correlation function is shown in Fig. \ref{BEC}.

{\bf MI.}-- For larger densities the phase coherence in the molecular state is destroyed
and a MI state with one bosonic molecule (i.e. a pair of fermions) appears. 
Due to strong phase fluctuations the second term in Eq. (\ref{homog2}) vanishes, and the correlation function 
becomes
\beq
C_{{\bf r},0}\propto - N^2\delta_{{\bf k},{\bf G}}+{\cal S}({\bf k})
\label{homog1}
\eeq
with the non-singular term
\beq
{\cal S}({\bf k})\approx\frac{J^2}{4\mu^2}(1-2 g_{\bf k}).
\label{eq2}
\eeq
Since the number of fermions $N$ is large in Eq. (\ref{homog1}), the contribution of ${\cal S}({\bf k})$
is negligible in time-of-flight experiments of the MI. 
Thus the MI state is characterized by sharp {\it dips} in contrast to the singular {\it peaks}
in the BEC, which appear on positions corresponding 
to the reciprocal lattice vectors ${\bf G}$, similar to a Bragg diffraction pattern.
However, in contrast to the latter, where the diffraction pattern is created by
light scattering on atoms of a crystal, here the pattern is created by the atomic state itself in the
expansion process of the cloud. 

\begin{figure}[h!!]
\begin{center}
\includegraphics[width=6cm]{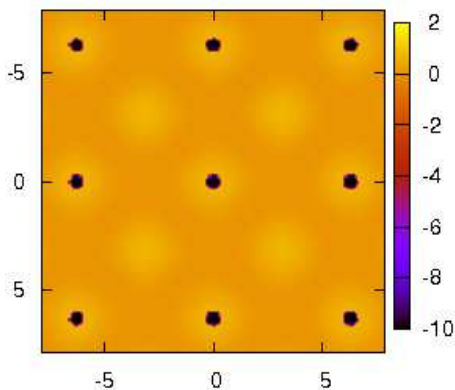}
\end{center}
\caption{Density-density correlation function as in Fig. \ref{BEC} but with $J=1$, $\mu=0.6$, $k_B T=0$.
The singular behavior in Fig. \ref{BEC} is replaced by dips here. This is due to the anticommuting nature 
of fermionic operators and thus reflects the fermionic nature of the MI state.}
\label{MI}
\end{figure}

Eq. (\ref{eq2}) describes the bosonic nature of the molecules in the MI state: There are hole 
excitations in the MI phase (particle excitations are suppressed by the hard-core nature of the bosons) and 
their dynamics contributes to the interference pattern via 
the factor $\propto g_{\bf k}$. A similar situation was observed experimentally in the case of bosonic Mott 
insulator in Ref. \cite{gerbier05}. The 2D plot of the correlation function is shown in Fig. \ref{MI}.


\section{Discussion}

The density-density correlation function of an expanding cloud provides
a clear picture of the state when it was still trapped by an 
optical lattice. It consists of four different contributions
in Eq. (\ref{noise2}), two of them are related to the fermionic nature of the atoms.
The second term leads to a fermionic dip in the density-density correlation function in the MI state.
A third term measures the condensate fraction in the case of condensed
bosonic molecules and together with  the singular fourth term leads to the sharp peaks.  
Even though the third term can compete with the second term, the singular term is the most relevant in the 
BEC phase.


The competing behavior of the fermionic dips and the condensate peaks
in the density-density correlation function are a result of the anticommuting
properties of the fermionic operators. This provides a simple concept to
distinguish different states in a cloud of attractively interacting fermions.
This behavior is rather different in a bosonic cloud, where all atoms contribute 
with the same sign to density-density correlation function because the bosonic operators
commute \cite{bloch}.

In conclusion, we have studied an expanding cloud of strongly interacting spin-1/2 fermions after
its release from an optical lattice.
The properties are described in terms of the density-density correlation function.
At lower densities a BEC is formed by the paired fermions, visible in the density-density correlation function
as sharp peaks. At higher densities a Mott-insulating phase appears, characterized by dips in the
density-density correlation function. This distinct behavior can be used in future experiments to distinguish
between these two phases in a gas of strongly attractive spin-1/2 fermions in an optical lattice.

\end{document}